\documentclass[twocolumn,prl,superscriptaddress,bibnotes,nofootinbib]{revtex4-1}
\usepackage{color}
\definecolor{red}{rgb}{0.75,0,0}
\definecolor{blue}{rgb}{0,0,0.75}
\definecolor{green}{rgb}{0,0.5,0}
\usepackage[pdftex, pdfborder={0 0 0}, colorlinks=true, linkcolor=red, urlcolor=blue]{hyperref}
\usepackage{amsmath}
\usepackage{amssymb}
\usepackage{lipsum}
\usepackage{bm}
\usepackage{graphicx}
\usepackage{verbatim}
\usepackage{mathtools}

\def\be{\begin{equation}}
\def\ee{\end{equation}}
\def\bea{\begin{eqnarray}}
\def\eea{\end{eqnarray}}

\def\besub{\begin{subequations}}
\def\eesub{\end{subequations}}

\def\bwd{\begin{widetext}}
\def\ewd{\end{widetext}}

\newcommand{\bsf}[1]{\textsf{\textbf{#1}}}

\newcommand{\SR}[1]{\textcolor{black}{#1}}

\newcommand{\AM}[1]{\textcolor{black}{#1}}

\begin{document}
\title{\SR{Oriented} Active Solids}
\author{Ananyo Maitra}
\email{nyomaitra07@gmail.com}
\affiliation{LPTMS, CNRS, Univ. Paris-Sud, Universit\'e Paris-Saclay, 91405 Orsay, France}
\author{Sriram Ramaswamy}
\email{sriram@iisc.ac.in}
\affiliation{Centre for Condensed Matter Theory, Department of Physics, Indian Institute of Science, 560 012 Bangalore, India}

\begin{abstract}
We present a complete analysis of the linearised dynamics of active \SR{\textit{solids}} with orientational order{, taking into account a hitherto overlooked consequence of rotation invariance}. Our predictions include the possibility of stable active solids with quasi-long-range order in two dimensions and long-range order in three dimensions, generic instability of the solid for one sign of active forcing, and {the instability of the orientationally ordered phase in momentum-conserved {systems} for large active forcing irrespective of its sign}. 

\end{abstract}

\maketitle
Biological systems and their artificial analogues are driven out of equilibrium by \SR{a direct and independent supply of free energy} to the individual constituent units. The nontrivial coupling of the constituent particles to this \SR{power} input \cite{SRJSTAT, LPDJSTAT} results in macroscopic stresses and currents in the system which  have been shown to be responsible for diverse phenomena ranging from coherent intracellular flows \cite{Ano_nuc_rot,Seb1, Gold}, to defect turbulence in living liquid crystals \cite{Dogic1}, to flocking in animal groups \cite{giardina}. Active hydrodynamics \cite{RMP}, which extends the traditional continuum dynamics of conserved and broken-symmetry variables by introducing terms that break only the time-reversal symmetry of the corresponding passive systems, offers a successful general framework to describe the macroscopic behaviour of such systems.

Whereas active versions of orientationally ordered \emph{fluids} have been studied extensively \cite{RMP, Sriram_rev, Ramaswamy_Toner_Tu, Curie_rep, Jacques_Nat}, active solids have received much less attention. In particular, spontaneous alignment in the solid phase, relevant for the dynamics of polarised tissues {on intermediate timescales}, has been somewhat neglected. 
Epithelial cellular monolayers \cite{philip}, and the cell \cite{shila1} itself, have been modelled as crosslinked active gels \cite{isol}, and models that couple phase-field crystals with the Toner-Tu equations for a polar flock have been constructed \cite{Lowen}. {Even the studies that consider anisotropic active gels \cite{shila1, Lowen, kopf-pismen}, however, ignore a symmetry-mandated coupling between orientation and strain in the free energy of distortion whose consequence is that one shear modulus of a uniaxial gel vanishes: true solids with uniaxial orientational order cannot exist in \SR{thermal} equilibrium. What \SR{about} an \textit{active} orientationally-ordered solid? To answer this question and more 
we present a complete hydrodynamic theory of active solids with uniaxial polar or apolar orientational order, in contact with a substrate or in a momentum conserving permeating fluid medium.} The former description is applicable to cell monolayers or tissues in an extracellular matrix while the latter describes the dynamics of active gels such as actomyosin at timescales that are small compared to the unbinding time of crosslinkers. We start with a description in which the polar or apolar order parameter is coupled to an \emph{isotropic} solid and, in the \emph{ordered} phase, eliminate the non-hydrodynamic director modes to obtain the effective equations for polar or apolar active gels. Our \SR{treatment applies to} the systems considered in \cite{Ewan} in the limit of infinite polymer relaxation time\SR{, 
and to the hydrodynamic limit of a Vicsek model with particles connected by springs }\cite{huepe}. 

{Here are our main results. }
(i) Active polar or apolar elastomers, when dynamically stable, have displacement fields whose variance is finite in three dimensions and diverges only logarithmically with system size in two dimensions. Translational order in such systems is quasi-long-range in two dimensions and long-range in three dimensions. This is in contrast to passive orientationally ordered elastomers which, lacking one shear modulus, can support only quasi-long-range order in three and short-range order in two dimensions. 
(ii) An active force proportional to $\propto{\bsf Q}\cdot\nabla\cdot{\bsf Q}$, where ${\bsf Q}$ is the apolar order parameter, always destabilises elastomers on a substrate when its magnitude is larger than the active force $\propto \nabla\cdot{\bsf Q}$, irrespective of the sign, in contrast to its stabilising role in incompressible active fluids on substrates. (iii) The dynamics of polar motile elastomers, at \emph{linear and nonlinear} order, is qualitatively distinct from that of a solid driven in an externally imposed direction \cite{Rangan} \SR{and escapes the latter's transverse buckling instability}. (iv) For bulk momentum-conserving systems, ignoring inertia, extensile (contractile) stresses destabilise positively (negatively) uniaxial elastomers. (v) Elastomeric gels are generically destabilised at high activities when a forcing rate given by the ratio of the coefficient of the active stress to the viscosity exceeds the passive orientational relaxation rate. (vi) Like active smectics \cite{Tapan, chen}, stable uniaxial active elastomers have finite concentration fluctuations and, in bulk momentum-conserving settings, a ``second sound'' mode whose speed is nonzero in all directions.  

We now demonstrate how we obtain these results. We start our discussion with apolar permanently crosslinked elastomers on a substrate. 
Our description involves the displacement field $\bar{{\bf u}}({\bf x}, t)$ about an isotropic reference state, the apolar order parameter field ${\bsf Q}({\bf x}, t)$, and the velocity field ${\bf v}({\bf x}, t)$. We define a linearised strain field ${\bsf W}={\bsf C}+\Phi{\bsf I}$, $W_{ij}=\partial_j\bar{u}_i+\partial_i\bar{u}_j$ about the isotropic state, where ${\bsf C}$ and $\Phi$ are the trace-removed and isotropic parts of ${\bsf W}$ and ${\bsf I}$ is the unit tensor.
Since the elastomer is permanently crosslinked there is no motion of structure without mass motion and the density field is slaved to {dilations of the structure, $\delta\rho/\rho=-\nabla\cdot\bar{{\bf u}}$,} implying that the dynamics of the density need not be explicitly considered. Further, since there is no permeation, the evolution equation for the displacement field is $\dot{\bar{{\bf u}}}={\bf v}$.
%

We assume a purely relaxational dynamics $\dot{{\bsf Q}}=-\Gamma_Q{\delta F}/{\delta {\bsf Q}}$, for the apolar order parameter ignoring all couplings to velocity \cite{FTNT0}. Here  
\begin{multline}
F=\int_{\bf x} \frac{\lambda}{2}\left(\Phi-\frac{s}{\lambda}\text{Tr}[{\bsf Q}^2]\right)^2+\mu\text{Tr}\left[{\bsf C}-\frac{t}{\mu}{\bsf Q}\right]^2+f_Q
\label{FQ}
\end{multline}
is the free-energy functional governing the dynamics in the absence of activity,  
with $f_Q=({r}/{2})\text{Tr}[{\bsf Q}^2]+({w}/{4})(\text{Tr}[{\bsf Q}^2])^2+(K/2)(\nabla{\bsf Q})^2$. \AM{The generic free-energy couplings $s$ and $t$ between the strain and the polarisation lead to the development of a strain anisotropy either parallel ($t>0$) or perpendicular to ($t<0$) ${\bsf Q}$. The absence of these couplings} \cite{shila1, Lowen} would have implied \AM{an invariance of the} uniaxial solid under {\it independent} rotations of the orientation and the elastic network, at least in the passive limit. Their inclusion {reduces} this symmetry to one under {\it joint} rotations of {the two}. 

The force balance equation is $\Gamma{\bf v}=\Gamma\dot{\bar{{\bf u}}}=-{\delta F}/{\delta \bar{{\bf u}}}+{\bf f}_a$, where the total active force is 
\begin{equation}
\label{apolfrc}
{\bf f}_a=\tilde{\zeta}_1\nabla\cdot{\bsf Q}+\tilde{\zeta}_2{\bsf Q}\cdot(\nabla\cdot{\bsf Q}).
\end{equation}
The first term in \eqref{apolfrc} is the divergence of the standard active stress while the second is an active force {density} that is allowed in all systems {not constrained by momentum conservation} and was first postulated in \cite{ano_apol}. 


Armed with these equations, we {write ${{\bsf Q}}=S({\bf n}{\bf n}-\bsf{I}/2)$ where the unit vector ${\bf n}$ and the scalar $S$ are the director and the magnitude of orientational order. We expand about the \emph{equilibrium} steady-state ${{\bsf Q}}={\bsf Q}^0$ with $S=S_0 =\sqrt{2r/w}$, ${\bf n} = {\bf n}_0=(1,0)$, $\Phi^0=({s}/{2\lambda})S_0^2$ and $\bsf{C} = \bsf{C}^0 =({t}/{\mu})\bsf{Q}^0$, obtained by minimising \eqref{FQ}, and {assess} how active stresses alter it}. Expanding the fields in small fluctuations about this steady state, $S=S_0+\delta S$, ${\bf n}=(\cos\theta, \sin\theta)\approx(1,\theta)$, $C_{ij}=C^0_{ij}+\delta C_{ij}$, and $\Phi=\Phi^0+\delta\Phi$, we find that the fluctuations of the apolar order parameter are slaved to the strain field as $\delta S=({2t}/{\bar{w}})\delta C_{xx}+({sS_0}/{\bar{w}})\delta\Phi
$
and
$
\theta=({\mu}/{S_0t})\delta C_{xy}$, where
$\bar{w}=S_0s^2/\lambda+t^2/\mu+r$. Note that due to the minimal couplings between ${\bsf Q}$ and ${\bsf C}$, director fluctuations are not {independently} soft and are instead slaved to the strain fluctuations \cite{Anderson}. 
Ignoring the free-energy coupling $\mu$ in \eqref{FQ} would have led to the incorrect conclusion that the transverse fluctuations of ${\bsf Q}$ have a relaxation rate that vanishes  {as wavelength $\to \infty$}.
\AM{
To analyse the displacement fluctuations about this orientationally ordered configuration, {we transform to displacement ${\bf u}$ and strain variables $\bm{\eta}$ relative to the anisotropic state from $\bar{{\bf u}}$ and ${\bsf C}$ and $\Phi$ which were defined relative to an isotropic reference space.} Upon integrating out the $\delta S$ fluctuations, the free energy \eqref{FQ} in terms of $\theta$ and ${\bf u}$ transforms to}
\begin{equation}
\label{fenrgeta}
F=\frac{1}{2}\int_{\bf x} B_1\eta_{xx}^2+B_2\eta_{yy}^2+B_3\eta_{xx}\eta_{yy}+B_4[\eta_{xy}-\beta(\theta-\Omega)]^2
\end{equation}
where $\eta_{ij}=(1/2)(\partial_i u_j+\partial_ju_i)$, the rotation angle $\Omega=(\partial_x u_y-\partial_yu_x)/2$. {$\beta \in [-1,1]$, whose sign is that of $t$, measures the degree of anisotropy of the solid. $\beta=0$ is the isotropic case, and we will refer to positive and negative anisotropy according to sgn$(\beta)$}. The standard analysis leading up to \eqref{fenrgeta} as well as the expression of $B_i$ and $\beta$ in terms of previously introduced coefficients are presented in full in the supplement \cite{supp}. Note that the angle \AM{$\theta$} appears with the rotation field of the solid and the shear strain and integrating it out leads to a vanishing of the shear modulus \cite{Olmsted, Ranjan, Xing, FTNT4}.

If the dynamics were controlled \AM{purely} by \AM{the} free-energy \eqref{fenrgeta} the hydrodynamic equations for $\dot{\bf u}$ would not contain terms of the form $\partial_y^2u_x\hat{x}$ and $\partial_x^2u_y\hat{y}$ {upon integrating out the fast $\theta$ field}. However, \AM{the} active force ${\bf f}_a =S_0(\tilde{\zeta}_1+{\tilde{\zeta}_2S_0}/{2})\partial_y\theta\hat{x}+S_0(\tilde{\zeta}_1-{\tilde{\zeta}_2S_0}/{2})\partial_x\theta\hat{y}$ yields terms of this form {even} when $\theta$ is eliminated in favour of $\Omega+\beta^{-1}\eta_{xy}$. 
Putting all of this together and rescaling time by the friction coefficient $\Gamma$, we obtain the hydrodynamic equations for the displacement fields:
\begin{subequations}
\label{fapol1}
\begin{gather}
\dot{u}_x=\nu_1\partial_x^2 u_x+\nu_2\partial_y^2u_x+\nu_3\partial_x\partial_yu_y\\
\dot{u}_y=\nu_4\partial_x^2 u_y+\nu_5\partial_y^2u_y+\nu_6\partial_x\partial_yu_x
\end{gather}
\end{subequations}
where $\nu_2$ and $\nu_4$ are purely active with $\nu_2=S_0(\beta^{-1}-1){(2\tilde{\zeta}_1+\tilde{\zeta}_2S_0)}/{4}$ and $\nu_4=S_0(\beta^{-1}+1){(2\tilde{\zeta}_1-\tilde{\zeta}_2S_0)}/{4}$. All other coefficients have both active and passive \AM{contributions} though $\nu_3-\nu_6$ is also purely active (see the supplement \cite{supp} for expressions for all the coefficients). Both $\nu_2$ and $\nu_4$ have to be positive for stability i.e. for $\beta>0$ stability requires $\tilde{\zeta}_1>|(\tilde{\zeta}_2S_0/2)|$ and for $\beta<0$, $\tilde{\zeta}_1<-|(\tilde{\zeta}_2S_0/2)|$. A negative $\tilde{\zeta}_1$ which denotes extensile stresses in this convention, destabilises a positively uniaxial solid with nematogens aligned along the deviatoric strain while a positive $\tilde{\zeta}_1$, denoting contractile active stresses, {destabilises} solid in which the nematogens align perpendicular to the local deviatoric strain anisotropy. \AM{This should be compared to the situation in active smectic{s \cite{Tapan} and in cholesterics \cite{Tapan_chol} or columnar systems.}. Interestingly, when $|\tilde{\zeta}_2|>2\tilde{\zeta}_1/S_0$, the orientationally ordered state is destabilised i.e., in contrast to incompressible active fluids on substrates, where, depending on the sign, it may  play a \emph{stabilising} role}, here $\tilde{\zeta}_2$ plays a \emph{destabilising} role, irrespective of its sign. This establishes our result (ii).

The presence of $\nu_2$ and $\nu_4$ ensures that the {variances} of both $u_x$ and $u_y$ fluctuations scale as $1/q^2$ in all directions when the elastomer is stable. In particular, $\langle u_x(q, t) u_x(-q,t)\rangle|_{q_x=0}\propto 1/(\nu_2q_y^2)$ and $\langle u_y(q,t) u_y(-q,t)\rangle|_{q_y=0}\propto 1/(\nu_4q_x^2)$, which are the directions in which the fluctuations would have been largest ($\propto 1/q^4$) in the absence of activity. This implies that {an active elastomer with apolar nematic order can support quasi-long-range translational order in two dimensions}. This conclusion remains correct even taking into account all symmetry allowed nonlinearities all of which are of the form $\nabla(\nabla {\bf u})^2$ and are irrelevant even in two dimensions in contrast to passive nematic elastomers where the absence of $\nu_2$ and $\nu_4$ terms render them relevant in {$d < 3$} \cite{Xing, Stenull}.

We now turn to \emph{polar} elastomers on substrates. \AM{The ordering, in this case, is characterised} by a polarisation vector ${\bf p}({\bf x}, t)$, instead of a rank-2 tensor. We again assume a purely relaxational dynamics for the polarisation field ignoring all advective and self-advective nonlinearities \cite{FTNT}
$
\dot{{\bf p}}=-\Gamma_p{\delta F_p}/{\delta {\bf p}},
$
where $F_p$ is of the form \eqref{FQ} with $\text{Tr}[{\bsf Q}^2]$ being replaced by ${p^2}$ and {$\bsf{Q}$ by $\bsf{D}$ with components}
$D_{ij}=p_ip_j-(1/2)p^2\delta_{ij}$ \cite{FTNT2}.
The force balance equation $\Gamma{\bf v}={\bf f}_p+\nabla\cdot\bm{\sigma}-\delta F/\delta\bar{{\bf u}}$ \AM{for polar systems} contain active propulsive forces ${\bf f}_p=(\upsilon\Gamma){\bf p}+(\upsilon_1\Gamma){\bf p}\cdot{\bsf W}$, in addition to the divergence of the active stress which we now take to be $\sigma_{ij}^a=\zeta_1D_{ij}+\zeta_2 p^2\delta_{ij}$ \cite{FTNT6}
%
We {assess} the effects of the active motilities and stresses on a state with ${\bf p}=p_0\hat{x}$, where $p_0^2=|r|/w$, $C_{ij}^0=t/\mu D_{ij}^0$, $\Phi^0=(s/\lambda)p_0^2$ and a steady velocity ${\bf v}=v_0\hat{x}=(\upsilon p_0+\upsilon_1p_0W_{xx}^0)\hat{x}$. Expanding the fields in small fluctuations, ${\bf p}\approx(p_0+\delta p)\hat{x}+p_0\theta\hat{y}$, $C_{ij}=C^0_{ij}+\delta C_{ij}$, and $\Phi=\Phi^0+\delta\Phi$, about the homogeneously oriented state \AM{we again find that both the magnitude  $p_0\delta p={s\delta\Phi+2t\delta C_{xx}}/{\bar{r}}$, where $\bar{r}=w+(2s^2/\lambda)+2(t^2/\mu)$, and the angular fluctuations $\theta=[{\mu}/{t p_0^2}]\delta C_{xy}$ of the polarisation field are slaved to the elastic deformations}. \SR{An analysis along the lines of the apolar case yields dynamical equations for the displacement} field ${\bf u}$ about the anisotropic stretched state. While the calculation for passive terms and terms originating from the active stress \SR{parallels} the apolar case (see the supplement \cite{supp} for details), the components of the propulsive force require more care. \AM{Projecting the propulsive force along and transverse to the polarisation, we obtain}
\begin{multline}
\hat{p}\cdot{\bf f}_p\stackrel{\text{lin}}{=}\Gamma[v_0+(\upsilon+\upsilon_1W_{xx}^0)\delta p+\upsilon_1p_0\delta W_{xx})\\=\Gamma v_0+a_1\partial_x u_x+a_2\partial_y u_y
\end{multline} 
where {$\stackrel{\text{lin}}{=}$ denotes equality to linear order in disturbances}, $\hat{p}$ is the unit vector in the direction of polarisation, {$a_1$ and $a_2$ are obtained by replacing $\delta p$ and $\delta W_{xx}$ by their values in terms of the displacement fields (see \cite{supp})} and 
$
\hat{p}_\perp\cdot{\bf f}_p\stackrel{\text{lin}}{=}\Gamma\upsilon_1p_0(\delta W_{xy}-2\theta C_{xx}^0),
$
where $\hat{p}_\perp$ is a unit vector perpendicular to $\hat{p}$. {Since $C_{xx}^0=(t/2\mu)p_0^2$ and $\theta=(\mu/tp_0^2)\delta C_{xy}=(\mu/tp_0^2)\delta W_{xy}$, $f_{a\perp}=0$. }This implies that there is no propulsive force in the direction transverse to the polarisation in a spontaneously symmetry broken polar elastomer. {Physically, since the only anisotropy in the system is due to the broken rotation symmetry, the anisotropy of ${\bsf W}$ aligns with ${\bf pp}$ on a fast timescale and at longer times, ${\bsf W}\cdot{\bf p}$ is purely parallel to ${\bf p}$.} {Next, projecting the velocity along and transverse to the direction of polarisation, we obtain $\hat{p}\cdot{\bf v}\stackrel{\text{lin}}{=}\dot{u}_x$ and $\hat{p}_\perp\cdot{\bf v}\stackrel{\text{lin}}{=}\dot{u}_y-v_0\theta$ where the final term appears due to the mean motion of the solid along $\hat{x}$ \cite{supp, ano_mem, RamTonPro}.}
{Therefore, upon transforming to a frame moving with the mean velocity of the solid $v_0\hat{x}$, }
\begin{subequations}
\begin{gather}
\dot{u}_x=a_1\partial_xu_x+a_2\partial_y u_y+b_1\partial_x^2 u_x+b_2\partial_y^2u_x+b_3\partial_x\partial_yu_y\\
\label{pseqn2}
\dot{u}_y=b_4\partial_x^2 u_y+b_5\partial_y^2u_y+b_6\partial_x\partial_yu_x
\end{gather}
\end{subequations}
where the coefficients $b_i$ are explicitly calculated in the supplement \cite{supp}. This yields the eigenfrequencies
\begin{equation}
\label{eigen1}
\omega_{\pm}=\begin{aligned}\begin{rcases*} \left(a_1+\frac{a_2b_6}{|b2-b5|}\right)q_x-ib_2q_y^2\\ -\frac{a_2b_6}{|b2-b5|}q_x-ib_5q_y^2\end{rcases*}q_x\ll q_y^2\\\begin{rcases*}a_1q_x-ib_1q_x^2-i\left(b_2+\frac{a_2}{a_1}b_6\right)q_y^2\\-ib_4q_x^2-i\left(b_5-\frac{a_2}{a_1}b_6\right)q_y^2\end{rcases*}q_x\gtrsim q_y^2\end{aligned}
\end{equation}
As in the nematic case, $b_2$ and $b_4$ are purely active $\propto\zeta_1$ and both of them are positive, implying that the polar solid is stable, only when $\zeta_1\beta<0$.
Motility introduces the possibility of a further instability when the magnitude of $\zeta_1$ is small: if the motility coefficients $a_2$ and $a_1$ have opposite signs then for small enough value of $b_2$, the system is destabilised. It is interesting to contrast \eqref{eigen1}  with the mode structure of an externally driven solid \cite{Rangan}. {There, terms proportional to $\partial_y u_x$ and $\partial_x u_y$ appear in the $\dot{u}_y$ equation even in the co-moving frame since the propulsion direction is externally determined (by gravity) and, as a result,} the eigenfrequencies are either wavelike at small wavevectors or have an instability with a growth rate $\propto q$. \AM{However, the direction of motion of crawling solids is chosen spontaneously forbidding this transverse buckling instability and implying that active solids may be more stable than both their passive counterparts and externally driven ones. This establishes our result (iii).}

As in apolar elastomers, the static structure factor of both $u_x$ and $u_y$ fluctuations scale as $1/q^2$ in all directions implying QLRO in two dimensions. However, this conclusion of the linear theory may be invalidated by marginally relevant nonlinearities. There are nine possible terms bilinear in $\nabla{\bf u}$ all of which are marginal in two dimensions. 
Unlike in {an externally driven solid} \cite{Rangan}, {where all of them may have independent values however, the coefficients of these nonlinearities are significantly constrained by rotational invariance in this case.} 
Despite this, a treatment including these nonlinearities is beyond the scope of this paper. We however note that the presence of two relevant nonlinearities $(\partial_x u)^2$ and $(\partial_y u)^2$, where $u$ is the displacement field of a smectic, in polar smectics does not destroy the ordered phase \cite{chen} 
when the two nonlinearities have opposite signs and believe that QLRO is possible even in polar solids in two dimensions for at least some range of parameter values.

While we have focussed on two dimensional elastomers, the three-dimensional case may be considered by changing $q_y$ and $u_y$ to $q_\perp$ and $u_\perp$ where $\perp$ denotes the two directions ($y,z$) transverse to the ordering direction $x$ and introducing an additional \AM{(passive)} elastic constant for $\eta_{yz}$ shears in the transverse plane in that case. While passive elastomers only have QLRO in three dimensions -- \SR{$\langle u_x^2 \rangle$ diverges} logarithmically as in smectics while $\langle |{\bf u}_\perp|^2 \rangle$ is finite as in columnar systems -- a calculation \SR{like that for two dimensions shows that} the static structure factor of displacement fluctuations diverges as $1/q^2$ in all directions in three-dimensional active elastomers implying long-ranged order. The absence of any relevant nonlinearity in three dimensions implies that this linear result is exact at large distances for both polar and apolar elastomers. \SR{Further \cite{supp}, active} \emph{biaxial} elastomers, whose passive analogues are softer than their uniaxial counterparts, with vanishing energy cost for all shear deformations and $u_x$, $u_y$ and $u_z$ all having smectic-like correlations, also have long-range order in three dimensions.

We now turn to elastomeric gels i.e. (dilute) elastomers frictionally coupled to an incompressible permeating 
\SR{fluid, with conserved total momentum of elastomer and fluid}. The velocity of the \AM{solid} is $\dot{{\bf u}}={\bf v}+{\bf v}_r$ where ${\bf v}$ is the joint velocity of the network and the fluid and ${\bf v}_r$ is the velocity of the network relative to the fluid. The {joint} velocity ${\bf v}$ obeys the force balance equation $\eta\nabla^2{\bf v}=\nabla\Pi-\tilde{\zeta}_1\nabla\cdot{\bsf Q}+\delta F/\delta{\bf u}$ where $\Pi$ is the pressure that enforces incompressibility, $\eta$ is the viscosity and $F$ is given by \eqref{fenrgeta} \cite{FTNT5}. Writing the force balance equation in terms of a stream function $\psi$, such that $v_x=\partial_y\psi$ and $v_y=-\partial_x\psi$, to take the incompressibility {constraint} into account, and defining $u_t={(q_yu_x-q_xu_y)}/{q}$ and $u_l={{\bf q}\cdot{\bf u}}/{q}$, we see that $\dot{u}_t=i|q|\psi+\mathcal{O}(q^2)$ and $\dot{u}_l\sim\mathcal{O}(q^2)$. 
\SR{The relaxation rate of $u_t$ fluctuations, which are the ones affected by the fluid, scale as $q^0$ to leading order in wavevector}. In a passive elastomer the $\mathcal{O}(q^0)$ part of the relaxation rate of $u_t$ has to vanish for perturbations either purely along the ordering direction i.e. with wavevectors ${\bf q}=q\hat{x}$ or purely transverse to it with wavevectors ${\bf q}=q\hat{y}$. However, in the active elastomer, the eigenfrequency in these directions \SR{is} non-zero 
:
\begin{equation}
\label{eigensimp}
\SR{\omega({\bf q} \to {\bf 0})=-i\frac{S_0\tilde{\zeta}_1}{2\eta}(\beta^{-1}\pm 1)}
\end{equation}
where the $+$ \SR{($-$) sign is realised for wavevectors  along $\hat{x}$ ($\hat{y}$)} (see the supplement \cite{supp} for the expression \SR{for all} wavevector directions). This implies that when $\beta\tilde{\zeta}_1<0$, the gel is unstable i.e. extensile (contractile) stresses destabilise positively (negatively) uniaxial elastomers, establishing our result (iv).
Note that while the relaxation rate of $u_t$ fluctuations is $\mathcal{O}(q^0)$ to leading order, its static structure factor scales as $1/q^2$ in all directions, when the gel is not destabilised, as can be checked by adding {momentum-conserving spatiotemporally white} noise to the force balance equation. 

We have, till now, assumed a fast relaxation of the angle field to its steady-state value determined by the displacement field. This may however be questionable in a gel -- an orientationally ordered fluid is generically unstable with \AM{a wavevector-independent} growth-rate \cite{Aditi1, RMP}. Therefore, when the {growth rate} of this generic instability is greater than the relaxation of the angle field to the value dictated by the local strain, the gel should be generically unstable \emph{irrespective} of the sign of the active stress. We now calculate this upper bound using the \emph{coupled} equations for the displacement fields and the angle field
$
\dot{\theta}=({1-\xi\cos2\phi})q^2\psi/2-\Gamma_\theta{\delta F}/{\delta\theta},
$
with $\xi$ being the flow-alignment parameter and $\Gamma_\theta$ being the angular relaxation rate.
We demonstrated \eqref{eigensimp} that a contractile gel with $\tilde{\zeta}_1>0$ is stable for $\beta>0$ in a description retaining only the displacement fields. Let us therefore take $\beta>0$, $\tilde{\zeta}_1>0$ (i.e., the case that we would predict to be stable based on a calculation purely in terms of displacement fields). {Also, for simplicity, let us take $\xi=0$ in which case, an orientationally ordered active fluid (not solid) is maximally unstable along $q_y$ i.e. pure splay. We calculate the $\mathcal{O}(q^0)$ part of the eigenfrequency of the coupled $u_t$ and $\theta$ dynamics in this direction} ($u_l$ dynamics does not contribute at this order):
\begin{multline}
\omega_{\pm}=\frac{i}{8\eta}\big[2S_0\tilde{\zeta}_1-B_4(1+4\beta^2\Gamma_\theta\eta)\pm \\\sqrt{\{2S_0\tilde{\zeta}_1-B_4(1+4\beta^2\Gamma_\theta\eta\}^2-32B_4(\beta-\beta^2)\Gamma_\theta S_0\tilde{\zeta}_1\eta}\big]
\end{multline}
{Therefore, when $S_0\tilde{\zeta}_1/\eta$ exceeds the relaxation rate of angular fluctuations in a passive nematic elastomeric gel, $B_4/2\eta+2B_4\beta^2\Gamma_\theta$, the orientational order is destabilised even in a positively uniaxial contractile solid since the angular fluctuations grow fast due to activity and can not relax to their preferred orientation dictated by the local strain. This yields an upper limit for the value of active stress above which a description purely in terms of the displacement fields is no longer possible and the orientationally ordered phase is destroyed irrespective of the sign of the active stress establishing our result (v).
}
%

We now leave the steady Stokesian regime, restore inertia $\rho_0\partial_t{\bf v}$ to the force balance equation and project the equations of motion of a momentum conserved stable active elastomer along and transverse to the wavevector direction, ignoring viscosity as is appropriate in the long wavelength limit. Along $q_x=0$ and $q_y=0$, the equations of motion decouple and the speed of the ``second" sound (i.e. the $u_t$ mode) vanishes in \textit{passive} elastomers as a consequence of vanishing restoring force for shear deformations. In contrast, in \textit{active} systems  the second sound speed along $q_x=0$ is $\sqrt{S_0\tilde{\zeta}_1(\beta^{-1}-1)/2\rho_0}$ while along $q_y=0$, it is $\sqrt{S_0\tilde{\zeta}_1(\beta^{-1}+1)/2\rho_0}$. 

Finally, since in elastomeric systems the density fluctuations are slaved to displacement fluctuations $\delta\rho/\rho=\nabla\cdot\bar{{\bf u}}$, and since $\langle uu\rangle \sim 1/q^2$ in all directions, the static structure factor of density fluctuations scales as $q^0$, i.e. there are no giant density fluctuations.

This concludes our discussion of active solids. Our theory applies to monolayers or tissues in an extracellular matrix and the dynamics of active gels such as actomyosin at timescales that are small compared to the unbinding time of crosslinkers.
Our results in this paper can be checked in simulations of Vicsek models in which the ``spins" are coupled by harmonic springs \cite{huepe} or in artificial active solids, for instance, composed of active chemotactic colloids \cite{suro}. 

\begin{acknowledgments}We thank M. E. Cates, M. C. Marchetti, E. Hemingway, S. Banerjee, P. Srivastava, S. Saha, S. Fielding, J. Toner, J.-F. Joanny and J. Prost for illuminating discussions. SR was supported by a J C Bose Fellowship of the SERB (India) and the Tata Education and Development Trust.\end{acknowledgments}


\begin{thebibliography}{}
\bibitem{FTNT0} Since all components of ${\bsf Q}$ will be shown to relax fast to values governed by the strain field,  all such velocity couplings are irrelevant in systems in contact with substrates.
\bibitem{FTNT4}{\eqref{fenrgeta} is analogous to the de Gennes model of smectics \cite{Chaikin} and the vanishing of the shear modulus is analogous to the vanishing of the tilt modulus in that case.}
\bibitem{FTNT2} Note that despite being a polar systems, the free energy is symmetric under ${\bf p}\to-{\bf p}$. The terms that break this symmetry are of the form $(\partial_i p_j+\partial_j p_i)C_{ij}$ \cite{shila1} appearing only at the next order in gradients and are thus irrelevant for the long wavelength physics we are interested in.
\end{thebibliography}
\end{document}